\documentclass[12pt]{article}
\usepackage{amssymb}
\usepackage{amsmath}
\usepackage{amsfonts}
\usepackage{amsthm}
\usepackage{bbm}
\usepackage{graphicx}
\usepackage{epstopdf}
\usepackage{color}
\usepackage[onehalfspacing]{setspace}
\usepackage{hyperref}
\usepackage{multirow}
\usepackage{float}
\usepackage{bm}
\usepackage{mathabx}
\usepackage{booktabs}
\usepackage{enumitem}
\usepackage{url}
\usepackage{dsfont}
\usepackage{algorithm}
\usepackage{algpseudocode}

\usepackage[flushleft]{threeparttable}

\usepackage[font=small]{caption}
\usepackage{tikz}
\usepackage{graphicx}
\usepackage{wrapfig}

\def\E{\mathbb{E}}
\def\P{\mathbb{P}}

\numberwithin{equation}{section}

\newtheorem{theorem}{Theorem}[section]
\newtheorem{lemma}{Lemma}[section]
\newtheorem{corollary}{Corollary}[section]
\newtheorem{remark}{Remark}[section]
\newtheorem{definition}{Definition}[section]

\theoremstyle{definition}
\newtheorem{assumption}{Assumption}%[section]

\theoremstyle{definition}
\newtheorem{example}{Example}[section]

\theoremstyle{definition}

\newcommand{\indep}{\perp\!\!\!\perp}

\usepackage{geometry}
\usepackage{color, hyperref}
\hypersetup{
    bookmarks=true,         % show bookmarks bar?
    unicode=false,          % non-Latin characters
    pdftoolbar=true,        %
    pdfmenubar=true,        %
    pdffitwindow=false,     % window fit to page when opened
    pdfstartview={FitH},    % fits the width of the page to the window
    pdftitle={My title},    % title
    pdfauthor={Author},     % author
    pdfsubject={Subject},   % subject of the document
    pdfcreator={Creator},   % creator of the document
    pdfproducer={Producer}, % producer of the document
    pdfkeywords={keyword1} {key2} {key3}, % list of keywords
    pdfnewwindow=true,      % links in new window
    colorlinks=false,       % false: boxed links; true: colored links
    linkcolor=black,          % color of internal links (change box color with linkbordercolor)
    citecolor=black,        % color of links to bibliography
    filecolor=black,      % color of file links
    urlcolor=black           % color of external links
}
\hypersetup{hidelinks}
\usepackage[authoryear,round]{natbib}
\setcitestyle{maxnames=2}

\title{
Unconfoundedness as an Information Constraint\thanks{
I thank Egshiglen Batbayar, Björn Höppner, Ruixuan Liu, Withney Newey, Bernard Salani\'e, 
Nico Thurow, and numerous seminar and conference participants for helpful comments and illuminating discussions. 
Support  from  the Deutsche Forschungsgemeinschaft (DFG, German Research Foundation) under Germany’s Excellence Strategy (EXC-2047/1 – 390685813) is gratefully acknowledged.
}
	}
		
\author{Christoph Breunig\thanks{Department of Economics, University of Bonn. Email: \url{cbreunig@uni-bonn.de}}}

\begin{document}
  \maketitle
\begin{abstract}
This paper provides an information-theoretic characterization of unconfoundedness within a general class of models allowing for endogenous selection. We impose a conditional mutual information (CMI) restriction requiring that the information about treatment contained in a potential outcome, conditional on covariates, does not exceed that contained in its latent heterogeneity. Under the maintained model conditions, this restriction is equivalent to a multiplicative factorization of treatment probabilities into separate functions of latent heterogeneity and covariates. Although the CMI restriction permits endogenous selection, sufficiently rich variation in treatment probabilities across covariate values makes it equivalent to unconfoundedness for the corresponding potential outcome. For binary treatments, any nonconstant propensity score suffices; for multiple treatments, a full-rank condition on the second-moment matrix of the propensity-score vector is sufficient. 
State-specific restrictions bound residual selection and imply exogeneity for individual treatment states under boundary-support conditions. Extensions cover continuous treatments and instrument exogeneity.
\end{abstract}
	\noindent{\footnotesize \noindent \textsc{Keywords:}  Causal inference, conditional mutual information, unconfoundedness, treatment probability variation, multiplicative factorization, latent heterogeneity. }
		\newpage
\section{Introduction}
Identification based on unconfoundedness requires potential outcomes to be independent of treatment conditional on observed covariates. Departures from this assumption are often described as selection on unobservables, without specifying how covariates and latent characteristics jointly determine treatment assignment. This paper uses a conditional mutual information restriction to characterize selection mechanisms that permit such departures. We show that sufficiently rich variation in treatment probabilities across covariate values makes this restriction equivalent to unconfoundedness for the corresponding potential outcome.
 
 We use conditional mutual information (CMI) to measure the dependence between treatment and a potential outcome after conditioning on covariates. Mutual information, introduced by \citet{shannon1948}, measures the statistical information shared between random variables, while its conditional analogue measures the information that remains shared after conditioning on additional variables; see, e.g., \citet{cover2006}. Rather than requiring the residual dependence to vanish, as under unconfoundedness, we require it to be no greater than the unconditional dependence between treatment and the latent heterogeneity entering the potential outcome. For example, in the evaluation of a training program, potential wages may remain informative about participation conditional on observed characteristics, provided that they are no more informative than latent wage heterogeneity. A key advantage of mutual information in our setting is that the symmetry of the associated information identities links the restriction directly to the treatment assignment mechanism.
 
Our first main result characterizes the selection mechanisms satisfying the conditional mutual information restriction (CMIR). We show that CMIR is equivalent to a multiplicative factorization of the conditional treatment probability into separate functions of latent outcome heterogeneity and covariates. This class includes unconfoundedness, selection on latent heterogeneity alone, and, with multiple treatment states, endogenous selection involving both covariates and latent heterogeneity. 
Our second main result establishes when variation in treatment probabilities across covariate values rules out residual confounding under CMIR. For binary treatment, any nonconstant propensity score suffices; for multiple treatments, a full-rank condition on the second-moment matrix of the propensity-score vector is sufficient. Under the maintained model conditions and the corresponding relevance condition, CMIR is therefore equivalent to unconfoundedness for the potential outcome under consideration.

We also consider information restrictions imposed only for a particular treatment state. Such restrictions may be useful when selection differs across treatment states, so that a restriction plausible for, say, non-participants is less plausible for participants. We show that a state-specific information restriction yields a corresponding factorization of the treatment probability. 
Under a boundary-support condition requiring the probability of the relevant treatment state to become arbitrarily close to one for some covariate values, the state-specific CMI restriction is equivalent to conditional independence between the potential outcome and the indicator for that treatment state, given covariates.

We develop several extensions of the framework. First, we allow for additional observed characteristics that may be related to latent heterogeneity. The independence condition is then imposed conditional on these characteristics, and the multiplicative characterization and equivalence with unconfoundedness extend under a corresponding conditional relevance condition. For continuous treatments, completeness of the conditional treatment distribution replaces the finite-dimensional rank condition. When interest is restricted to the average potential outcome at a given treatment value, pointwise relevance and suitable regularity conditions suffice for identification without full unconfoundedness. Finally, we extend the framework to instrumental-variable models. Maintaining the conventional exclusion restriction, we formulate CMI restrictions that relax instrument exogeneity and establish conditions under which they are equivalent to outcome-side or joint instrument exogeneity.\\

\textit{Related Literature:} 
Conditional independence between treatment and potential outcomes underlies propensity-score methods for binary, multivalued, and continuous treatments \citep{rosenbaumrubin1983,imbens2000,hiranoimbens2004}. 
We instead allow residual dependence and characterize when a mutual information restriction is sufficient to recover unconfoundedness.

Information-theoretic quantities have been used in causal inference to measure causal influence \citep{AyPolani2008,JanzingEtAl2013}, as balancing criteria for matching in observational studies \citep{sun2016}, and to bound causal effects under unmeasured confounding \citep{JungKang2026}.
In economics and econometrics, entropy and related measures have been used to study dependence, aggregation, and distributional restrictions \citep{Maasoumi1993}. Mutual information measures information acquisition in models of rational inattention \citep{Sims2003,MatejkaMcKay2015} and provides a basis for tests of unconditional and conditional independence \citep{AiEtAl2024}. Kullback--Leibler divergence also underlies misspecification-robust inference in partially identified models \citep{KaidoMolinari2026}.

The literature on sensitivity to unobserved confounding and partial identification studies how causal conclusions change when unconfoundedness is relaxed, for example by relating selection on unobservables to selection on observables \citep{AltonjiElderTaber2005,Oster2019}, bounding omitted-variable bias \citep{ChernozhukovEtAl2022}, or characterizing identified sets under nonparametric departures from conditional independence \citep{MastenPoirier2018}. We instead characterize a class of selection mechanisms permitting departures from unconfoundedness and establish when variation in treatment probabilities across covariate values rules out residual confounding within this class.
\\

The remainder of the paper is organized as follows. Section \ref{sec:model} introduces mutual information in the potential-outcomes framework and formulates the conditional mutual information restriction. Section \ref{sec:im} establishes its multiplicative characterization and derives conditions under which it is equivalent to unconfoundedness for a given potential outcome. Section \ref{sec:ext} extends the framework to additional covariates that may be related to latent heterogeneity, continuous treatments, and instrumental-variable models, where we consider information-theoretic relaxations of instrument exogeneity. Auxiliary results and proofs are collected in Appendix \ref{appendix:main:proofs}. A simulation study in Appendix \ref{sec:simulation} examines the finite-sample performance of a rank test for covariate relevance.

\section{Setup and the CMI Restriction}\label{sec:model}
This section introduces the conditional mutual information restriction (CMIR) in the potential outcome framework. We interpret the restriction in terms of treatment selection and illustrate selection mechanisms that satisfy it.
\subsection{Setup}\label{sec:setup}
Let $D$ denote a multivalued treatment with support $\mathcal D=\{0,1,\ldots,K\}$, and let $\{Y_d\}_{d\in\mathcal D}$ denote the associated potential outcomes. The observed outcome is $Y=\sum_{d\in\mathcal D}\mathbbm{1}_{\{D=d\}}Y_d$, where $\mathbbm{1}_{\{\cdot\}}$ denotes the indicator function. In addition, a vector of covariates $X$ is available. For each $d\in\mathcal D$, we represent the potential outcome as
\begin{align}\label{mod:outcome}
Y_d=m_d(X,U_d),\qquad U_d\indep X,
\end{align}
where $U_d$ captures latent outcome heterogeneity. This representation can be obtained by taking $U_d$ to be a latent rank in the conditional distribution of $Y_d$ given $X$.\footnote{\label{fn:latent:rank}If $Y_d\mid X$ is continuously distributed, the conditional probability integral transform yields a latent rank $U_d=F_{Y_d\mid X}(Y_d\mid X)$, where $F_{Y_d\mid X}(y\mid x)=\mathbb P(Y_d\leq y\mid X=x)$, that is uniformly distributed and independent of $X$. Moreover, $Y_d$ can be recovered almost surely from the corresponding conditional quantile function. 
If the conditional distribution has mass points, the randomized probability integral transform assigns a uniform rank within the probability interval associated with each mass point; see \citet[Theorem 1]{Brockwell2007}.
Multivariate outcomes can be handled using the Rosenblatt transform or its randomized extension; see \cite{Rosenblatt1952,Brockwell2007}.}
Throughout this paper, we refer to $U_d$ as latent heterogeneity and formulate our information restrictions relative to the chosen latent representation. Section \ref{sec:ext} extends the framework to $U_d\indep X\mid W$ for additional covariates $W$ and to continuously distributed treatments.

For the identification of causal parameters such as average treatment effects, it is common to impose unconfoundedness for each potential outcome, that is, $Y_d\indep D\mid X$ for $d\in\mathcal D$. Under model \eqref{mod:outcome}, $U_d\indep D\mid X$ is sufficient for this condition and, if $m_d(x,\cdot)$ is injective for almost every $x$, also necessary. Rather than imposing this independence, we allow treatment to depend on latent heterogeneity through the potential outcome, conditional on covariates. Specifically, for $d\in\mathcal D$, we assume
\begin{align}\label{mod:treatment}
U_d\indep D\mid (Y_d,X).
\end{align}
Condition \eqref{mod:treatment} requires that latent heterogeneity provide no additional information about treatment once the potential outcome and covariates are given. It holds automatically when $U_d$ can be recovered from $(Y_d,X)$, as with the conditional-rank representation for continuously distributed outcomes in footnote \ref{fn:latent:rank}, or whenever $m_d(x,\cdot)$ is injective for almost every $x$. For discrete outcomes, several latent values can produce the same outcome, and the condition requires that treatment probabilities do not distinguish between these values conditional on $X$. The randomized rank construction in footnote \ref{fn:latent:rank} satisfies this condition when the auxiliary randomization is independent of $(Y_d,D,X)$.

The condition \eqref{mod:treatment} also accommodates treatment decisions based on the potential outcome, covariates, and additional factors, provided that these factors contain no further information about latent heterogeneity conditional on $(Y_d,X)$. For example, in the Roy model $D=\mathbbm{1}_{\{Y_1-Y_0>0\}}$, workers choose the sector offering higher potential earnings. Given $(Y_1,X)$, treatment is determined by earnings in the alternative sector. A sufficient condition for \eqref{mod:treatment} with $d=1$ is therefore $Y_0\indep U_1\mid(Y_1,X)$.

		\subsection{Mutual Information}\label{sec:mi}
	Mutual information, introduced by \citet{shannon1948}, is the Kullback--Leibler divergence between the joint distribution of two random variables and the product of their marginal distributions.
\begin{definition}[Mutual Information]
The mutual information of a pair of random variables $V$ and $R$ is defined by
\begin{align*}
I(V;R)=\E\left[\log\left(\frac{p(V,R)}{p(V)p(R)}\right)\right],
\end{align*}
and the conditional mutual information of $(V,R)$ given $X$ is defined by
\begin{align*}
I(V;R\mid X)=\E\left[\log\left(\frac{p(V,R\mid X)}{p(V\mid X)p(R\mid X)}\right)\right],
\end{align*}
where $p$ denotes the relevant joint, marginal, or conditional density or probability mass function.
\end{definition}

Mutual information measures the information that observing one random variable provides about another. In general, the mutual information between two random variables can be either larger or smaller than their conditional mutual information given another variable. For a given $d\in\mathcal D$, we impose the conditional mutual information restriction
\begin{align*}
\mathrm{CMIR}(d):\quad I(Y_d;D\mid X)\leq I(U_d;D).
\end{align*}
The restriction has the following interpretation:
\begin{center}
\textit{The potential outcome, conditional on covariates, is no more informative about treatment than latent heterogeneity.}
\end{center}
For example, suppose that $Y_d$ denotes potential wages and $D$ indicates participation in a training program. The restriction allows potential wages to remain informative about participation after conditioning on covariates, provided that this information does not exceed the unconditional information about participation contained in latent wage heterogeneity. When potential wages are continuously distributed conditional on covariates, the rank normalization in footnote \ref{fn:latent:rank} identifies $U_d$ with the individual's rank in this conditional distribution. The right-hand side then measures how informative this relative wage position under treatment $d$ is about actual participation.

Under model \eqref{mod:outcome}--\eqref{mod:treatment}, the potential outcome preserves all information about treatment contained in latent heterogeneity conditional on $X$. Thus, $I(Y_d;D\mid X)=I(U_d;D\mid X)$; see Lemma \ref{lem:end} in Appendix \ref{appendix:main:proofs}. Consequently, $\mathrm{CMIR}(d)$ is equivalent to
\begin{align*}
I(U_d;D\mid X)\leq I(U_d;D).
\end{align*}
Thus, $\mathrm{CMIR}(d)$ permits dependence between treatment and latent heterogeneity, but requires that conditioning on $X$ does not increase the mutual information between them.
In the training example, participation may depend on latent wage heterogeneity, such as unobserved motivation, ability, or productivity, provided that conditioning on covariates does not make this heterogeneity more informative about participation.

\begin{remark}[Mutual information as a measure of selection]
Mutual information is not the only nonparametric measure of dependence. Alternatives include total variation, Hellinger and $\chi^2$ distances, maximal and distance correlation, and kernel-based measures.\footnote{Kernel-based measures are typically formulated in reproducing kernel Hilbert spaces (RKHSs); prominent examples are the Hilbert--Schmidt independence criterion (HSIC) and kernel conditional independence measures.}
Mutual information has a likelihood-ratio interpretation, is invariant to one-to-one transformations, and does not require a parametric selection equation. Of particular importance for our analysis is the symmetry of interaction information, which allows us to characterize the CMI restriction through a multiplicative factorization of the treatment assignment mechanism.
\end{remark}

\begin{example}[Unconfoundedness]\label{ex:uc}
A standard assumption for identifying causal effects is unconfoundedness for each potential outcome, $Y_d\indep D\mid X$ for $d\in\mathcal D$; see, e.g., \cite{rosenbaum1984,imbens2004}. This condition is equivalent to $I(Y_d;D\mid X)=0$ for each $d\in\mathcal D$ and therefore implies $\mathrm{CMIR}(d)$, since $I(U_d;D)\geq0$. Unconfoundedness thus provides the benchmark case in which the left-hand side of the CMI restriction vanishes, whereas CMIR also permits $I(Y_d;D\mid X)>0$.
\end{example}
The following examples show that CMIR admits endogenous selection through latent heterogeneity, potential outcomes, or both observed and latent characteristics. Section \ref{sec:im} establishes when variation in treatment probabilities across covariate values rules out such departures from unconfoundedness.

\subsubsection{Selection on Latent Heterogeneity}
We consider selection mechanisms in which treatment assignment may depend directly on latent heterogeneity. The \textit{interaction information} of $(U_d,D,X)$ is defined by
\begin{align*}
I(U_d;D;X):=I(U_d;D)-I(U_d;D\mid X).
\end{align*}
Recall that, by Lemma \ref{lem:end}, $I(U_d;D\mid X)=I(Y_d;D\mid X)$ under model \eqref{mod:outcome}--\eqref{mod:treatment}.
Hence, $\mathrm{CMIR}(d)$ is equivalent to $I(U_d;D;X)\geq 0$. A key feature of interaction information is its symmetry under permutations of its arguments, first established by \cite{mcgill1954}; see also \cite{cover2006}. In particular, we have
\begin{align}
\label{eq:perm}
I(U_d;D;X)=I(X;U_d;D)=I(D;X;U_d).
\end{align}

\begin{example}[Selection on latent heterogeneity]\label{ex:lh}
For a given $d\in\mathcal D$, suppose that treatment may depend on latent heterogeneity but is independent of covariates conditional on $U_d$, that is, $D\indep X\mid U_d$. Since $I(D;X\mid U_d)=0$, we have $I(D;X;U_d)=I(D;X)\geq 0$. By symmetry of interaction information, $I(U_d;D;X)\geq 0$, and hence $\mathrm{CMIR}(d)$ holds. Section 4.1 extends this argument to settings in which $U_d$ and $X$ may be dependent, provided that they are independent conditional on additional covariates.
\end{example}

\subsubsection{Selection on potential outcomes}
We next consider selection on potential outcomes, where treatment may depend directly on the potential outcome, but  covariates provide no additional information about treatment once the potential outcome is known. The following result characterizes when such selection is compatible with the CMI restriction.
\begin{lemma}\label{lem:send:sol}
In model \eqref{mod:outcome}--\eqref{mod:treatment}, for any $d\in\mathcal D$, suppose that $D\indep X\mid Y_d$. Then
\begin{align}\label{ineq:end:out}
I(Y_d;D\mid U_d)\leq I(X;D)
\end{align}
is equivalent to   $\mathrm{CMIR}(d)$.
\end{lemma}
Related conditional independence restrictions have been considered in the literature on selection and nonignorable nonresponse; see, for example, \citet{Manski1994}, \citet{Chen2001}, and \citet{TangLittleRaghunathan2003}. \citet{d'hault2010} develops a nonparametric identification approach based explicitly on independence between an instrument and a binary selection indicator conditional on the outcome. As in Section \ref{sec:im:cond}, the restriction considered here can be relaxed by requiring it to hold only conditional on additional covariates. 
Under selection on potential outcomes, $\mathrm{CMIR}(d)$ holds precisely when the information about treatment contained in potential outcomes beyond latent heterogeneity does not exceed the information contained in covariates.

\begin{example}[Selection on potential outcomes]
Let $X$ and $U_d$ be independent, finite-valued random variables, with $U_d$ nondegenerate. Suppose that $Y_d=m_d(X,U_d)$, where $m_d$ is injective on their joint support. Treatment is determined by $D=\phi_d(Y_d)$, where $\phi_d$ assigns a distinct treatment state to each possible outcome. 
Conditional on $U_d$, the potential outcome and covariates determine each other, so $I(Y_d;D\mid U_d)=I(X;D\mid U_d)$. Since treatment reveals $X$ and $X\indep U_d$, this equals $I(X;D)$, which verifies \eqref{ineq:end:out}. 
Likewise, conditional on $X$, the potential outcome and latent heterogeneity determine each other and thus $I(Y_d;D\mid X)=I(U_d;D)>0$.
Hence, $\mathrm{CMIR}(d)$ holds by Lemma \ref{lem:send:sol}.
\end{example}

\subsubsection{Within-Treatment Independence}
Models of sectoral self-selection allow observed characteristics and unobserved heterogeneity to jointly determine sector choice; see \citet{HeckmanSedlacek1985,HeckmanHonore1990}. Conditioning on sector choice may therefore induce dependence between $X$ and $U_d$, even though they are independent in the population. The following lemma establishes that $\mathrm{CMIR}(d)$ is equivalent to the absence of this within-treatment dependence.

\begin{lemma}\label{lem:lat:bal}
In model \eqref{mod:outcome}--\eqref{mod:treatment}, $\mathrm{CMIR}(d)$ is equivalent to $U_d\indep X\mid D$. Whenever $\mathrm{CMIR}(d)$ holds, it holds with equality; if $D$ is also binary, then either $U_d\indep D$ or $X\indep D$.
\end{lemma}
Lemma \ref{lem:lat:bal} provides a complementary interpretation of the information restriction in terms of the composition of treatment groups: within each group, the distribution of latent heterogeneity does not vary with covariates. Under the maintained model conditions, CMIR holds with equality whenever it is satisfied, although both sides may be positive. With multiple treatment states, this condition allows treatment assignment to depend on both $X$ and $U_d$. With binary treatment, however, the lemma implies either $U_d\indep D$ or $X\indep D$, so that assignment depends on at most one of these components.

\begin{example}[Selection on observables and unobservables]\label{ex:within}
Consider a program offering two forms of training. Participation depends on latent earnings heterogeneity, while assignment between training options depends on covariates.
For a given $d\in\mathcal D$, let $X,U_d,\eta_X,\eta_U$ be mutually independent standard normal variables and suppose that $Y_d=\alpha_d+\beta_dX+U_d$, with $\beta_d\neq0$. Treatment is determined by
\begin{align*}
D=\mathbbm{1}_{\{U_d+\eta_U>0\}}\bigl(1+\mathbbm{1}_{\{X+\eta_X>0\}}\bigr).
\end{align*}
Individuals participate when latent earnings heterogeneity exceeds a random threshold; participants receive training option $1$ or $2$ according to their covariates and an independent assignment shock. Since participation and assignment depend on separate, independent components, conditioning on treatment preserves independence between covariates and latent heterogeneity. Hence, $U_d\indep X\mid D$, and Lemma \ref{lem:lat:bal} implies that $\mathrm{CMIR}(d)$ holds. Nevertheless, participation selects on latent earnings heterogeneity, so unconfoundedness fails.
\end{example}

\section{Characterization and Implications of CMIR}\label{sec:im}
This section characterizes the implications of CMIR for the treatment assignment mechanism and establishes conditions under which the restriction is equivalent to unconfoundedness for the corresponding potential outcome. We also consider treatment-state-specific versions of CMIR and derive corresponding exogeneity results.
\subsection{Factorization of the Treatment Assignment Mechanism}
We first provide an equivalent representation of the CMI restriction in terms of a multiplicative structure of the conditional treatment probabilities. This factorization provides the link between the information restriction and unconfoundedness: sufficiently rich variation in treatment probabilities across covariate values rules out dependence of treatment on latent heterogeneity conditional on covariates.
Let $\mathcal U_d:=\operatorname{supp}(U_d)$ and $\mathcal X:=\operatorname{supp}(X)$.

 \begin{theorem}\label{thm:mult}
In model \eqref{mod:outcome}--\eqref{mod:treatment}, $\mathrm{CMIR}(d)$ is equivalent, for each $d\in\mathcal D$, to the existence of nonnegative measurable functions $g_{d,d'}:\mathcal U_d\to\mathbb R_+$ and $h_{d'}:\mathcal X\to\mathbb R_+$ with
\begin{align*}
\P(D=d'\mid U_d,X)=g_{d,d'}(U_d)h_{d'}(X)\quad \text{for all }d'\in\mathcal D.
\end{align*}
\end{theorem}
 
Theorem \ref{thm:mult} shows that CMIR restricts the treatment assignment mechanism to a multiplicative structure, with separate components for latent heterogeneity and covariates. This characterization encompasses unconfoundedness for $Y_d$ as well as the selection mechanisms discussed in Section \ref{sec:mi}. Moreover, the factorization can be normalized so that $h_{d'}(X)=\P(D=d'\mid X)$ for every $d'\in\mathcal D$. Thus, the covariate component can be chosen to be the same across potential outcomes.

The proof of Theorem \ref{thm:mult} combines the symmetry of interaction information in \eqref{eq:perm} with the normalization $U_d\indep X$ to derive the multiplicative factorization. A related multiplicative characterization is established by \citet{RennerMaurer2002}, who require correlation freeness, meaning that a mutual-information inequality holds for every joint input distribution with the conditional output distribution held fixed. This uniform requirement is not suitable for our potential-outcomes setting, where CMIR concerns only the given joint distribution.\\

We next consider a mutual-information restriction imposed only for a particular treatment state, which yields a factorization of the corresponding conditional treatment probability. To formulate this restriction, we introduce a treatment-state-specific version of mutual information, motivated by the literature on information measures associated with particular realizations of a random variable \citep{DeWeeseMeister1999,Butts2003}. The unconditional version below corresponds to the \textit{specific surprise} of \citet{DeWeeseMeister1999}.

\begin{definition}[Specific Mutual Information]
Suppose that $R$ is discrete and let $r$ satisfy $\P(R=r)>0$. The $r$-specific mutual information between $V$ and $R$ is defined as
\begin{align*}
I_r(V;R)=\E\left[\left.\log\left(\frac{p(V\mid R=r)}{p(V)}\right)\right|R=r\right].
\end{align*}
Similarly, the $r$-specific conditional mutual information between $V$ and $R$ given $X$ is defined as
\begin{align*}
I_r(V;R\mid X)=\E\left[\left.\log\left(\frac{p(V\mid R=r,X)}{p(V\mid X)}\right)\right|R=r\right].
\end{align*}
\end{definition}
The quantity $I_r(V;R)$ measures how informative the event $R=r$ is about $V$, and equals the Kullback--Leibler divergence between the distribution of $V$ conditional on $R=r$ and its marginal distribution. Similarly, $I_r(V;R\mid X)$ measures the corresponding information beyond $X$, averaging over the distribution of $X$ conditional on $R=r$. There is no general ordering between $I_r(V;R)$ and $I_r(V;R\mid X)$.

For a given treatment state $d'\in\mathcal D$, we impose the state-specific restriction
\begin{align*}
\mathrm{CMIR}(d',d):\quad I_{d'}(Y_d;D\mid X)\leq I_{d'}(U_d;D).
\end{align*}
The left-hand side measures the information about the potential outcome gained by learning that an individual receives treatment $d'$, when their covariates
\begin{wrapfigure}{r}{0.45\textwidth}
\centering
\resizebox{\linewidth}{!}{%
\begin{tikzpicture}[font=\normalsize,line width=0.45pt]
\filldraw[fill=black!4] (0,0) ellipse [x radius=4.6cm,y radius=2.65cm];
\filldraw[fill=black!10] (0,-0.46) ellipse [x radius=3.6cm,y radius=1.82cm];
\filldraw[fill=black!17] (0,-0.95) ellipse [x radius=2.25cm,y radius=0.85cm];
\node at (0,1.92) {$\mathrm{CMIR}(d',d)$ given $d'\in\mathcal D$};
\node at (0,0.64) {$\mathrm{CMIR}(d)$};
\node at (0,-0.95) {$Y_d\indep D\mid X$};
\end{tikzpicture}%
}
\label{fig:cmir:nesting}
\end{wrapfigure}
 are already known.  The right-hand side measures the information about latent heterogeneity gained from the same observation, without conditioning on covariates. The restriction requires the former to be no greater than the latter. Imposing $\mathrm{CMIR}(d',d)$ for every $d'\in\mathcal D$ is equivalent to $\mathrm{CMIR}(d)$, as established by Theorem \ref{thm:mult} and Corollary \ref{coro:mult:d} below.

The usual mutual and conditional mutual information decompose as
\begin{align*}
I(V;R)=\sum_r \P(R=r) I_r(V;R),
\qquad I(V;R\mid X)=\sum_r \P(R=r) I_r(V;R\mid X);
\end{align*}
see, e.g., \cite{DeWeeseMeister1999}.
Moreover, by Bayes' rule, the treatment-state-specific interaction information satisfies the symmetry identity $I_r(V;R)-I_r(V;R\mid X)=
I_r(X;R)-I_r(X;R\mid V)$. 
\begin{corollary}\label{coro:mult:d}
In model \eqref{mod:outcome}--\eqref{mod:treatment}, $\mathrm{CMIR}(d',d)$ is equivalent, for each $d,d'\in\mathcal D$, to the existence of nonnegative measurable functions $g_{d,d'}:\mathcal U_d\to\mathbb R_+$ and $h_{d'}:\mathcal X\to\mathbb R_+$ with
\begin{align*}
\P(D=d'\mid U_d,X)=g_{d,d'}(U_d)h_{d'}(X).
\end{align*}
\end{corollary}

As an example, consider a job-training program where $Y_0$ denotes potential earnings without training, $X$ covariates such as education or distance to the training center, and $U_0$ latent wage heterogeneity. Covariates and latent heterogeneity may both affect participation. Suppose, however, that among non-participants, the distribution of latent wage heterogeneity does not vary with covariates. For example, non-participants living closer to the training center have the same distribution of latent wage heterogeneity as those living farther away. This selection pattern satisfies
\begin{align*}
I_0(Y_0;D\mid X)\leq I_0(U_0;D).
\end{align*}
The corresponding restriction for participants need not hold: participants living farther from the training center may differ systematically in latent wage heterogeneity from those living closer. The state-specific restriction therefore allows the selection mechanism to differ across treatment groups.

Moreover, $\mathrm{CMIR}(0,0)$ is weaker than the conventional exogeneity condition $Y_0\indep D\mid X$. While exogeneity requires treatment to provide no information about untreated potential outcomes beyond $X$, $\mathrm{CMIR}(0,0)$ allows non-participation to provide such information, provided that it does not exceed what non-participation reveals about latent heterogeneity. For example, among individuals with the same education and distance to the training center, non-participation may signal lower latent earning ability and hence lower potential earnings without training.

\subsection{Equivalence to Unconfoundedness}
We now provide conditions under which the CMI restriction is equivalent to unconfoundedness for the corresponding potential outcome. We first consider multivalued treatments and then specialize to binary treatment. Let $p(X)=\bigl(\P(D=0\mid X),\ldots,\P(D=K\mid X)\bigr)^\top$ denote the propensity-score vector.

\begin{assumption}[Multinomial Covariate Relevance]\label{A:mult:relevance}
The matrix $\E[p(X)p(X)^\top]$ has full rank $K+1$.
\end{assumption}

Assumption \ref{A:mult:relevance} excludes any nonzero linear combination of the propensity scores that vanishes almost surely. For binary treatment, it reduces to $\operatorname{Var}\bigl(\P(D=1\mid X)\bigr)>0$, requiring a nonconstant propensity score. The condition is distinct from overlap, which restricts how close treatment probabilities can be to zero or one; neither condition implies the other. Since the rank condition concerns only the joint distribution of treatment and covariates, it can be assessed using rank-testing methods. Appendix \ref{sec:simulation} examines the finite-sample performance of the rank test of \citet{ChenFang2019} and illustrates the bias of inverse-probability weighting under residual confounding. While Examples \ref{ex:lh}--\ref{ex:within} illustrate that CMIR permits different forms of selection on unobservables, the next result establishes that the full-rank condition rules out the resulting departures from unconfoundedness.

\begin{theorem}\label{thm:uc:mult}
In model \eqref{mod:outcome}--\eqref{mod:treatment}, suppose that Assumption \ref{A:mult:relevance} holds. Then $\mathrm{CMIR}(d)$ is equivalent to $Y_d\indep D\mid X$ for each $d\in\mathcal D$. Consequently, imposing $\mathrm{CMIR}(d)$ for all $d\in\mathcal D$ is equivalent to unconfoundedness for each potential outcome separately.
\end{theorem}

Under the full-rank condition in Theorem \ref{thm:uc:mult}, the multiplicative assignment structure implies that treatment probabilities do not depend on latent outcome heterogeneity conditional on covariates. We now specialize to binary treatment, for which Assumption \ref{A:mult:relevance} requires only a nonconstant propensity score. Although the next result follows directly from Theorem \ref{thm:uc:mult}, we provide a separate proof exploiting the fact that the two conditional treatment probabilities sum to one.

\begin{corollary}\label{thm:uc}
In model \eqref{mod:outcome}--\eqref{mod:treatment} with binary treatment, suppose that $\P(D=1\mid X)$ is not almost surely constant. Then $\mathrm{CMIR}(d)$ is equivalent to $Y_d\indep D\mid X$ for each $d\in\{0,1\}$.
\end{corollary}

The multiplicative representation must hold for both treatment states, whose conditional probabilities sum to one. Any variation in the propensity score across covariate values therefore forces the normalized latent factors to equal one, eliminating dependence of treatment on latent heterogeneity conditional on $X$.\footnote{The conventional joint condition $(Y_0,Y_1)\indep D\mid X$ admits an analogous characterization. Under $(U_0,U_1)\indep X$ and $(U_0,U_1)\indep D\mid(Y_0,Y_1,X)$, the same argument shows that, if the propensity score is not almost surely constant, $I((Y_0,Y_1);D\mid X)\leq I((U_0,U_1);D)$ is equivalent to joint unconfoundedness.}

We finally consider the case in which the CMI restriction is imposed only for a particular treatment state. Since the restriction now pertains to a single treatment state rather than the treatment variable as a whole, the rank condition above is no longer sufficient to eliminate latent selection. Instead, the following result bounds the remaining dependence using the support of the corresponding treatment probability and establishes state-specific exogeneity when this probability can become arbitrarily close to one.

\begin{theorem}\label{thm:uc:state}
In model \eqref{mod:outcome}--\eqref{mod:treatment}, fix $d,d'\in\mathcal D$ and suppose that, for some $\kappa\in(0,1]$,
\begin{align*}
\P\bigl(\P(D=d'\mid X)>\kappa-\varepsilon\bigr)>0
\qquad\text{for every }\varepsilon>0.
\end{align*}
If $\mathrm{CMIR}(d',d)$ holds, then
\begin{align*}
0\leq I_{d'}(Y_d;D\mid X)=I_{d'}(U_d;D)\leq\log(1/\kappa).
\end{align*}
In particular, if $\kappa=1$, then $\mathrm{CMIR}(d',d)$ is equivalent to $\P(D=d'\mid Y_d,X)=\P(D=d'\mid X)$.
\end{theorem}

Larger admissible values of $\kappa$ yield tighter bounds on the state-specific information quantities. Under $\mathrm{CMIR}(d',d)$, the support of the treatment probability therefore limits the information associated with selection into state $d'$. For $\kappa<1$, both quantities may be positive; at $\kappa=1$, they must vanish.

The case $\kappa=1$ recovers the boundary-support condition: the probability of state $d'$ must approach one on sets of positive probability. This can hold without the probability attaining one and is compatible with pointwise, but not uniform, overlap. With finitely supported covariates, it requires treatment state $d'$ to occur with certainty in at least one covariate group. For binary treatment, boundary support requires propensity scores approaching one when $d'=1$ and zero when $d'=0$. This condition is related to the identification-at-infinity argument of \cite{Heckman1990}.

Like the rank condition, the support condition concerns only the observed treatment and covariate distribution. With finitely supported covariates, one-sided exact binomial tests within covariate groups, adjusted for multiple testing, can assess whether any treatment probability exceeds a given $\kappa<1$. Under CMIR, rejection supports the corresponding information bound. The same procedure can use thresholds $\kappa_n<1$ with $\kappa_n\to1$ to assess increasingly stringent support conditions. For continuously distributed covariates, the same tests can be applied within prespecified covariate regions, since a regional treatment probability exceeding $\kappa$ implies that the propensity score exceeds $\kappa$ on a set of positive probability. The choice of regions affects power; inference using increasingly fine partitions and thresholds approaching one is beyond the scope of this paper.

With binary treatment and $\kappa=1$, independence between the potential outcome and the indicator for either treatment state implies $Y_d\indep D\mid X$. Hence, under boundary support for a given $d'\in\{0,1\}$, $\mathrm{CMIR}(d',d)$ is equivalent to unconfoundedness for $Y_d$. Imposing the restriction for both potential outcomes yields unconfoundedness for each potential outcome separately.

\begin{example}[Average treatment effect on the treated (ATT)]\label{exmp:att}
Let us illustrate the result for identification of the ATT, $\E[Y_1-Y_0\mid D=1]$, in the binary-treatment case. Considered on its own, $\mathrm{CMIR}(0,0)$ permits departures from exogeneity of untreated potential outcomes. For example, non-participation in a training program may reveal information about untreated earnings beyond covariates, provided that this information does not exceed what non-participation reveals about latent wage heterogeneity. Suppose that $\mathrm{CMIR}(0,0)$ holds and that $\P\left(\P(D=1\mid X)<\varepsilon\right)>0$ for every $\varepsilon>0$. This support condition requires covariate regions of positive probability in which assignment to the control group is arbitrarily likely.

Under the boundary-support condition, Theorem \ref{thm:uc:state} implies that $\mathrm{CMIR}(0,0)$ is sufficient for $Y_0\indep D\mid X$. Adding the overlap condition $\P(D=1\mid X)\leq 1-\eta$ for some $\eta>0$ therefore identifies the ATT. The two support conditions concern opposite tails of the propensity score: boundary support requires values arbitrarily close to zero, whereas overlap requires the propensity score to remain bounded away from one.
\end{example}

\begin{example}[State-specific exogeneity in a Roy model]\label{ex:roy:state}
Following the multisector Roy tradition \citep{HeckmanSedlacek1985,Magnac1991}, we consider a three-sector example illustrating which aspects of endogenous sorting a state-specific information restriction excludes. Let $X,U,V$ be mutually independent standard normal variables. Potential log earnings are $Y_0=\alpha U+\gamma|U|$, $Y_1=X+(\alpha+1)U+V$, and $Y_2=X+(\alpha-1)U+V$, where $\alpha>1$ and $\gamma\in[0,1]$. The sign of $U$ determines comparative advantage between sectors $1$ and $2$, while $X$ and $V$ shift earnings in both sectors.

Workers choose $D=\operatorname*{arg\,max}_{j\in\{0,1,2\}}Y_j$. Thus, $D=0$ if $X+V+(1-\gamma)|U|\leq0$; otherwise, workers choose sector $1$ when $U>0$ and sector $2$ when $U<0$. Taking $U_0=U$ satisfies the maintained conditions for $Y_0$, since $U\indep X$ and $Y_0$ is strictly increasing in $U$. The choice probability is $\P(D=0\mid U,X)=\Phi(-X-(1-\gamma)|U|)$. For $\gamma<1$, workers with larger $|U|$ are more likely to leave sector $0$. Moreover, $p_0(x)=\E[\Phi(-x-(1-\gamma)|U|)]$ converges to one as $x\to-\infty$, establishing boundary support.

The restriction $\mathrm{CMIR}(0,0)$ requires sector-$0$ choice to provide no more information about $Y_0$ conditional on $X$ than about $U$ unconditionally. By Theorem \ref{thm:uc:state}, it holds here precisely when $\gamma=1$, eliminating selection into sector $0$ based on $|U|$ while preserving endogenous sorting between sectors $1$ and $2$. Theorem \ref{thm:uc:state} establishes the sector-specific exogeneity implication without requiring these particular earnings functions or distributions.
\end{example}

\section{Extensions}\label{sec:ext}
This section extends the framework to additional covariates related to latent heterogeneity, continuous treatments, and instrumental-variable models.
\subsection{Conditioning on Additional Covariates}\label{sec:im:cond}
We now extend the framework by allowing the information restriction to condition on additional observed characteristics $W$. This is useful when the relationship between treatment selection and latent heterogeneity differs across observed subpopulations. 
For example, in program evaluation, the relationship between latent earnings heterogeneity and treatment participation may differ across regions, cohorts, industries, or local labor-market conditions.

We consider the conditional model
\begin{align}
\label{mod:outcome:cond}
Y_d=m_d(X,W,U_d),\qquad U_d\indep X\mid W.
\end{align}
As in the baseline model, the independence condition can be satisfied by taking $U_d$ to be a conditional rank of $Y_d$ given $(X,W)$. Thus, the role of $W$ is not to justify the latent representation, but to allow the information restriction to accommodate selection mechanisms that differ across observed characteristics. We further maintain
\begin{align}
\label{mod:treatment:cond}
U_d\indep D\mid(Y_d,X,W).
\end{align}
As in the baseline model, this condition holds automatically if $U_d$ is recoverable from $(Y_d,X,W)$. More generally, the sufficient conditions discussed in Section \ref{sec:setup} continue to apply after conditioning additionally on $W$.

We next consider an extension of the CMI restriction that conditions on additional covariates $W$. Specifically, let
\begin{align*}
\mathrm{CMIR}_W(d): \quad I(Y_d;D\mid X,W) \leq I(U_d;D\mid W).
\end{align*}
This restriction requires that, given $W$, the information about treatment contained in the potential outcome after controlling for $X$ does not exceed that contained in latent heterogeneity. The benchmark cases discussed above extend directly to this setting: unconfoundedness for the corresponding potential outcome is now imposed conditional on $(X,W)$, while selection on latent heterogeneity is allowed to vary with $W$. Likewise, the multiplicative characterization of the treatment assignment mechanism extends by allowing both components to vary with $W$. The next result formalizes this extension.

\begin{theorem}\label{thm:mult:cond}
In model \eqref{mod:outcome:cond}--\eqref{mod:treatment:cond}, $\mathrm{CMIR}_W(d)$ is equivalent, for each $d\in\mathcal D$, to the existence of nonnegative measurable functions $g_{d,d'}:\mathcal U_d\times\mathcal W\to\mathbb R_+$ and $h_{d'}:\mathcal X\times\mathcal W\to\mathbb R_+$ with
\begin{align*}
\P(D=d'\mid U_d,X,W)=g_{d,d'}(U_d,W)h_{d'}(X,W)\quad\text{for every }d'\in\mathcal D.
\end{align*}
\end{theorem}

We next characterize when the conditional CMI restriction is equivalent to unconfoundedness for the corresponding potential outcome given $(X,W)$. Let $p(X,W)=\bigl(\P(D=0\mid X,W),\ldots,\P(D=K\mid X,W)\bigr)^\top$ denote the vector of conditional treatment probabilities. The following result follows from Theorem \ref{thm:uc:mult}, applied conditionally on $W$, with the conditional rank condition replacing Assumption \ref{A:mult:relevance}. Its proof is therefore omitted.

\begin{corollary}\label{cor:uc:mult:cond}
In model \eqref{mod:outcome:cond}--\eqref{mod:treatment:cond}, suppose that $\operatorname{rank}(\E[p(X,W)p(X,W)^\top\mid W])=K+1$ almost surely. Then $\mathrm{CMIR}_W(d)$ is equivalent to $Y_d\indep D\mid(X,W)$ for each $d\in\mathcal D$.
\end{corollary}
Corollary \ref{cor:uc:mult:cond} implies that imposing $\mathrm{CMIR}_W(d)$ for every $d\in\mathcal D$ is equivalent to  unconfoundedness for each potential outcome conditional on $(X,W)$. The conditional rank condition requires sufficiently rich variation in the treatment probabilities with $X$, conditional on $W=w$, for almost every $w$.  For binary treatment, it reduces to $\operatorname{Var}\left(\P(D=1\mid X,W)\mid W=w\right)>0$
for almost every $w$, so that the propensity score must vary with $X$ conditional on $W=w$.
When $W$ has finite support, the conditional rank condition can be assessed using rank tests within each group. For continuously distributed $W$, conditional relevance could be assessed using sieve estimates of conditional moment matrices, but developing valid inference for this extension is beyond the scope of this paper.
% Valid inference would additionally require control of approximation bias and simultaneous uncertainty across conditioning values.

\subsection{Continuous Treatments}\label{subsec:uc:continuous}
We now extend the preceding results to a continuously distributed treatment. Let $D\in\mathcal D\subseteq\mathbb R^r$, and suppose that the conditional distributions of $D$ given $X$ and $(U_d,X)$ admit densities with respect to Lebesgue measure, denoted by $f_{D\mid X}$ and $f_{D\mid U_d,X}$, respectively. Let $\{Y_t:t\in\mathcal D\}$ be a jointly measurable potential-outcome process, and assume $Y=Y_D$. Throughout this subsection, we assume that the information quantities appearing in our restrictions are finite. For finite-valued treatments, this holds automatically.

The multiplicative characterization in Theorem \ref{thm:mult} extends directly to continuous treatments by replacing conditional treatment probabilities with conditional densities. For each $d\in\mathcal D$, $\mathrm{CMIR}(d)$ is equivalent to the existence of nonnegative measurable functions $g_d:\mathcal D\times\mathcal U_d\to\mathbb R_+$ and $h:\mathcal D\times\mathcal X\to\mathbb R_+$ such that
\begin{align}\label{eq:cont:gps:factorization}
f_{D\mid U_d,X}(t\mid u,x)=g_d(t,u)h(t,x)
\end{align}
for almost every $(t,u,x)$. The argument follows directly from the proof of Theorem \ref{thm:mult}, using $I(U_d;X\mid D)=0$, or equivalently $U_d\indep X\mid D$. The continuous-treatment analogue of the rank condition in Assumption \ref{A:mult:relevance} is a completeness condition on the conditional distribution of $D$ given $X$.

\begin{assumption}[Continuous-Treatment Covariate Relevance]\label{A:cont:relevance}
For every measurable function $\varphi:\mathcal D\to\mathbb R$ with $\E[|\varphi(D)|]<\infty$, $\E[\varphi(D)\mid X]=0$ implies $\varphi(D)=0$ almost surely.
\end{assumption}

The completeness requirement can be weakened to bounded completeness whenever the normalized deviations introduced below are bounded; see \citet{DHaultfoeuille2011} and \citet{Andrews2017} for sufficient conditions and examples of boundedly complete distributions. At the end of this subsection, we further weaken completeness to a functional relevance condition that is sufficient for identification of average potential outcomes.

\begin{corollary}\label{coro:uc:mult:cont}
In model \eqref{mod:outcome}--\eqref{mod:treatment} with a continuously distributed treatment,  suppose that Assumption \ref{A:cont:relevance} holds. Then $\mathrm{CMIR}(d)$ is equivalent to $Y_d\indep D\mid X$ for each $d\in\mathcal D$.
\end{corollary}

Corollary \ref{coro:uc:mult:cont} shows that completeness plays the same role for continuous treatments as the full-rank condition for multiple treatments. Sufficiently rich variation in the conditional distribution of $D$ given $X$ eliminates residual selection on latent heterogeneity, so that $\mathrm{CMIR}(d)$ is equivalent to unconfoundedness for the corresponding potential outcome. We next show that pointwise relevance, together with suitable regularity conditions, suffices to identify $\E[Y_d\mid X]$, and hence also $\E[Y_d]$, without requiring completeness.

\paragraph{Conditional mean independence.}

Rather than requiring full conditional independence between $Y_d$ and $D$ given $X$, identification here relies on a conditional mean restriction at the treatment value $d$. This restriction follows from the structure of the treatment assignment mechanism. In particular, it is sufficient that the normalized latent component of the multiplicative representation equal one at $d$.
Below, let $\mathcal H$ be a linear space of measurable functions $\varphi:\mathcal D\to\mathbb R$ satisfying $\E[|\varphi(D)|]<\infty$.
\begin{assumption}[Pointwise Covariate Relevance]\label{A:pointwise:relevance}
For a given $d\in\mathcal D$ and every $\varphi\in\mathcal H$, $\E[\varphi(D)\mid X]=0$ implies $\varphi(d)=0$.
\end{assumption}

Assumption \ref{A:pointwise:relevance} ensures that any two functions in $\mathcal H$ with the same conditional expectation given $X$ agree at $d$, even if they differ elsewhere. Thus, the conditional expectation identifies the value of a function at $d$ without necessarily identifying the entire function. This is closely related to the identification of linear functionals under noninjective conditional expectation operators studied by \cite{SeveriniTripathi2006}.\footnote{Define the conditional expectation operator by $(T\varphi)(x)=\E[\varphi(D)\mid X=x]$. Assumption \ref{A:pointwise:relevance} requires point evaluation at $d$ to vanish on the null space $\mathcal N(T)$. If $\mathcal H$ is a reproducing kernel Hilbert space with kernel $k$ and $T:\mathcal H\to L_X^2:=\{\phi:E[\phi^2(X)]<\infty\}$ is well-defined and bounded, this condition is equivalent to $k_d\in\mathcal N(T)^\perp$, where $k_d(\cdot)=k(\cdot,d)$ represents evaluation at $d$.} Pointwise relevance concerns only the observed distribution of $(D,X)$ for a given function space $\mathcal H$. When $\mathcal H$ is finite-dimensional, pointwise relevance can be expressed as a matrix range condition and tested under suitable regularity conditions.
The pointwise relevance condition in Assumption \ref{A:pointwise:relevance} can hold even when completeness in Assumption \ref{A:cont:relevance} fails.\footnote{Let  $\mathcal H=\{\varphi:\varphi(t)=a+b(t-d),\ a,b\in\mathbb R\}$ and $D=d+\sigma(X)\varepsilon$, where $\E[\varepsilon\mid X]=0$, $\E|D|<\infty$, and $\P(D\neq d)>0$. Then $\E[D\mid X]=d$, so $(T\varphi)(X)=a$ and $\mathcal N(T)=\{\varphi:\varphi(t)=b(t-d),\ b\in\mathbb R\}$. Completeness fails because $\E[D-d\mid X]=0$ while $D-d$ is nonzero with positive probability. Nevertheless, every $\varphi\in\mathcal N(T)$ satisfies $\varphi(d)=0$, and hence pointwise relevance at $d$ holds.} Below, $\|\cdot\|$ denotes the Euclidean norm.

\begin{assumption}[Lipschitz Continuity at $d$]\label{A:outcome:lipschitz}
There exist constants $L,\delta>0$ such that, almost surely, $|Y_t-Y_d|\leq L\|t-d\|$ for all $t\in\mathcal D$ satisfying $\|t-d\|<\delta$.
\end{assumption}
The Lipschitz condition limits how much individual potential outcomes can change as treatment moves away from $d$, with a common bound across individuals. It allows nonlinear responses and heterogeneous treatment effects. Since $Y=Y_D$ the condition implies that $|Y-Y_d|\leq L\|D-d\|$ whenever $\|D-d\|<\delta$, controlling the approximation error as treatment approaches $d$. Related Lipschitz restrictions on individual potential outcomes appear in \citet{deChaisemartinEtAl2024}, who allow an individual-specific Lipschitz bound subject to a conditional moment restriction. Here, we impose a common bound only locally around the treatment value of interest. 
For the identification result below, define $\widetilde g_d(t,u)=g_d(t,u)/\E[g_d(t,U_d)]$ whenever $0<\E[g_d(t,U_d)]<\infty$, and set $\widetilde g_d(t,u)=1$ otherwise.

\begin{theorem}\label{thm:apo:pointwise}
In model \eqref{mod:outcome}--\eqref{mod:treatment} with a continuously distributed treatment, suppose for a given $d\in\mathcal D$ that $Y_d$ is bounded, $\widetilde g_d(\cdot,u)-1\in\mathcal H$ for almost every $u$, and $\mathrm{CMIR}(d)$ holds along with Assumptions \ref{A:pointwise:relevance} and \ref{A:outcome:lipschitz}.
 Further, suppose that $f_{D\mid X}(d\mid \cdot)$ is bounded away from zero, that $\widetilde g_d(\cdot,u)$ is continuous at $d$ for almost every $u$, and that $f_{D\mid X}(\cdot\mid x)$ is continuous at $d$ for almost every $x$. Then $\E[Y_d\mid X]=\E[Y\mid D=d,X]$.
\end{theorem}

Theorem \ref{thm:apo:pointwise} provides a structural justification for identifying the conditional average potential outcome at a given treatment value. Rather than requiring covariate variation to eliminate latent selection throughout the treatment distribution, pointwise relevance ensures that the normalized latent factor equals one at $d$. Latent characteristics may therefore affect the conditional treatment density at other treatment values while leaving it unchanged at $d$. Completeness and full unconfoundedness may fail, while, under the stated regularity conditions, $\E[Y_d\mid X]$ remains identified from $\E[Y\mid D=d,X]$.

\begin{example}[Failure of pointwise relevance]
Let $U\in\{-1,1\}$ with equal probabilities and $U\indep X$. For $t\in(-1,1)$, let $Y_t=\mu(X)+\tau t+\sigma(X)U$ and $Y=Y_D$, where $\mu$ and $\sigma$ are bounded and $\sigma(X)>0$. Thus, $U_d=U$ for every $d$. Suppose
\begin{align*}
f_{D\mid U,X}(t\mid u,x)=(1+\alpha ut)f_0(t\mid x),
\qquad 0<|\alpha|<1,
\end{align*}
where $f_0(\cdot\mid x)$ is a positive, continuous, symmetric density on $(-1,1)$ that varies with $x$,  with $f_0(d\mid x)$ uniformly positive over $x$. Thus, $X$ determines the baseline treatment distribution, while $U$ shifts probability toward larger or smaller treatment values.

Since $U$ is determined by $(Y_d,X)$, condition \eqref{mod:treatment} holds, and the multiplicative factorization implies $\mathrm{CMIR}(d)$. Averaging over $U$ gives $f_{D\mid X}(t\mid x)=f_0(t\mid x)$ and hence $\E[D\mid X]=0$. Therefore, if $\varphi(t)=t\in\mathcal H$, pointwise relevance fails for every $d\neq0$ because $\E[\varphi(D)\mid X]=0$ while $\varphi(d)\neq0$. Moreover, $\E[U\mid D=d,X]=\alpha d$, so $\E[Y_d\mid X]=\mu(X)+\tau d$, whereas $\E[Y\mid D=d,X]=\mu(X)+\tau d+\alpha d\,\sigma(X)$. Thus, despite bounded potential outcomes and Lipschitz continuity, CMIR alone does not imply equality of these conditional means when pointwise relevance fails.
\end{example}

\subsection{Instrumental Variables}\label{subsec:mi:iv}
We now apply the mutual-information framework to instrument exogeneity while maintaining the conventional exclusion restriction. Let $\mathcal Z=\{0,\ldots,L\}$ denote the support of an instrument $Z$, and suppose that potential outcomes satisfy $Y_{d,z}=Y_d$ for all $(d,z)$. In potential-outcomes formulations of IV identification, instrument validity typically also restricts the dependence between $Z$ and the potential outcomes. We focus first on the outcome-side exogeneity condition $Y_d\indep Z\mid X$. This condition may fail, for instance, when an eligibility rule or encouragement has no direct effect on outcomes and hence satisfies exclusion, but its assignment still depends on latent characteristics such as motivation or productivity that also affect potential outcomes.
For each $d\in\mathcal D$, suppose that
\begin{align}
U_d\indep Z\mid(Y_d,X).
\label{mod:instrument}
\end{align}
Instead of imposing $Y_d\indep Z\mid X$, consider
\begin{align}
I(Y_d;Z\mid X)\leq I(U_d;Z).
\label{eq:iv:cmir}
\end{align}
This restriction permits the instrument to be related to latent outcome heterogeneity, while requiring the potential outcome conditional on $X$ to be no more informative about the instrument than latent heterogeneity.

The multiplicative characterization in Theorem \ref{thm:mult} applies directly with $Z$ in place of $D$. Consequently, sufficiently rich variation in the conditional instrument probabilities eliminates the latent component of this representation, just as in the treatment-assignment setting. Let $p_Z(X)=\big(\P(Z=0\mid X),\ldots,\P(Z=L\mid X)\big)^\top$.

\begin{corollary}\label{cor:iv:exogeneity}
In model \eqref{mod:outcome}, suppose that exclusion $Y_{d,z}=Y_d$ and condition \eqref{mod:instrument} hold, and that $\operatorname{rank}\bigl(\E[p_Z(X)p_Z(X)^\top]\bigr)=L+1$. Then condition \eqref{eq:iv:cmir} is equivalent to $Y_d\indep Z\mid X$ for each $d\in\mathcal D$.
\end{corollary}
The result follows directly from Theorem \ref{thm:uc:mult}, with $Z$ in place of $D$ and the conditional instrument probabilities replacing the conditional treatment probabilities, so its proof is omitted. The rank condition requires sufficiently rich variation in instrument probabilities across covariate values, whereas conventional IV relevance requires the instrument to generate variation in the endogenous treatment.

\paragraph{Local average treatment effects.}
The preceding result concerns exogeneity with respect to each potential outcome separately. The standard local average treatment effect (LATE) framework instead imposes joint exogeneity of the instrument with respect to potential outcomes and potential treatment choices, conditional on covariates \citep{ImbensAngrist1994,Abadie2003}. For binary $D,Z\in\{0,1\}$, let $Y_d=m_d(X,U_d)$ and $D_z=\psi(z,X,V)$ for $d,z\in\{0,1\}$, with $(U_0,U_1,V)\indep X$. The observed treatment and outcome are $D=D_Z$ and $Y=Y_D$.

Collect the potential outcomes and potential treatments in $\Xi=(Y_0,Y_1,D_0,D_1)$, and let $S$ denote their joint latent heterogeneity,  with
\begin{align}\label{eq:iv:joint-representation}
\Xi=r(X,S), \qquad S\indep X,
\end{align}
for an appropriate measurable function $r$. One possible choice is $S=(U_0,U_1,V)$, provided that this vector is independent of $X$.
We further impose
\begin{align}\label{eq:iv:joint-sufficiency}
S\indep Z\mid(\Xi,X).
\end{align}
Thus, conditional on the potential outcomes, potential treatment choices, and covariates, instrument assignment contains no additional information about the latent heterogeneity determining these variables. This condition holds automatically if $S$ can be recovered from $(\Xi,X)$.

Analogously to \eqref{eq:iv:cmir}, consider the joint CMI restriction
\begin{align}\label{eq:iv:joint:cmir}
I(\Xi;Z\mid X)\leq I(S;Z).
\end{align}
This restriction allows instrument assignment to depend on latent heterogeneity, while bounding the information about the instrument contained in the joint vector of potential outcomes and potential treatment choices. 
For example, when eligibility for a training subsidy instruments participation, the restriction allows eligibility to be related to potential earnings and participation choices, provided that these variables, conditional on covariates, are no more informative about eligibility than their joint latent heterogeneity is unconditionally.
%Under \eqref{eq:iv:joint-representation}--\eqref{eq:iv:joint-sufficiency}, the preceding arguments apply directly to $(\Xi,S)$.

\begin{example}\label{exmp:iv:cmir}
Let joint latent heterogeneity be represented by a scalar $S\in\{0,1\}$ satisfying $S\indep X$ and $0<\P(S=1)<1$.
For some measurable function $\mu$ and $\tau>0$, let $D_z=zS$ for $z\in\{0,1\}$ and $Y_d=\mu(X)+S+d\tau$ for $ d\in\{0,1\}$.
Conditional on $X$, the vector $\Xi=(Y_0,Y_1,D_0,D_1)$ is a one-to-one function of $S$.
Let $\P(Z=1\mid S,X)=q_S$,
where $q_0,q_1\in(0,1)$ and $q_1\neq q_0$.
Joint instrument exogeneity fails because $\Xi\not\indep Z\mid X$ but \eqref{eq:iv:joint:cmir} holds since $S$ is recoverable from $(\Xi,X)$ and $X\indep(S,Z)$. 
The Wald estimand does not have the conditional LATE interpretation of \citet[Proposition 5.3]{Abadie2003}:
\begin{align*}
\frac{\E[Y\mid Z=1,X=x]-\E[Y\mid Z=0,X=x]}
{\E[D\mid Z=1,X=x]-\E[D\mid Z=0,X=x]}
=
\tau+1-\frac{\pi_0}{\pi_1},
\end{align*}
where $\pi_z=\P(S=1\mid Z=z)$. By contrast, $\E[Y_1-Y_0\mid D_1>D_0,X=x]=\tau$.
\end{example}

In the preceding example, $S\indep X$ and $\P(Z=1\mid S,X)=\P(Z=1\mid S)$ imply that the instrument propensity $\P(Z=1\mid X)$ is constant. The next result shows that, under \eqref{eq:iv:joint-representation}--\eqref{eq:iv:joint-sufficiency}, a nonconstant instrument propensity makes the joint CMI restriction equivalent to joint instrument exogeneity.

\begin{corollary}\label{cor:iv:late}
Under \eqref{eq:iv:joint-representation}--\eqref{eq:iv:joint-sufficiency}, suppose that $\P(Z=1\mid X)$ is not almost surely constant. Then the joint CMI restriction \eqref{eq:iv:joint:cmir} is equivalent to $\Xi\indep Z\mid X$.
\end{corollary}

The proof follows directly from the argument of Theorem \ref{thm:uc:mult}, applied to the joint representation above, and is therefore omitted. The joint CMI restriction allows instrument assignment to depend on latent heterogeneity underlying both potential outcomes and potential treatments and, unlike the outcome-specific restrictions, also restricts its relationship with the potential treatment choices $(D_0,D_1)$. Corollary \ref{cor:iv:late} shows that a nonconstant instrument propensity makes this restriction equivalent to joint instrument exogeneity. Together with exclusion, monotonicity, overlap, and a nonzero first stage, joint instrument exogeneity yields the standard conditional LATE interpretation of the Wald estimand.

\section{Conclusion}

This paper studies when an information restriction permitting selection on latent outcome heterogeneity can justify identification based on unconfoundedness. We impose a conditional mutual information restriction that, on its own, allows treatment to remain related to potential outcomes after conditioning on covariates. We characterize the selection mechanisms satisfying this restriction and show that sufficiently rich variation in the conditional treatment distribution across covariate values makes CMIR equivalent to unconfoundedness for the corresponding potential outcome. For multivalued treatments, the rank condition concerns only the joint distribution of treatment and covariates and can be assessed using rank-testing methods.

When the relevance conditions hold, residual confounding necessarily violates the information restriction; when they fail, endogenous selection may remain compatible with it. State-specific restrictions provide further identification results. For the ATT, the information restriction need only concern non-participation and the untreated potential outcome. Although this restriction alone permits selection on untreated potential outcomes, combining it with boundary support implies their exogeneity, and overlap then ensures identification. For continuous treatments, pointwise relevance and suitable regularity conditions suffice to identify an average potential outcome without full unconfoundedness.

\appendix
\section{Auxiliary Results and Proofs}\label{appendix:main:proofs}
This appendix collects auxiliary lemmas and proofs of the main results.
\subsection{Auxiliary Lemmas}\label{appendix:lemmas}

\begin{lemma}\label{lem:end}
In model \eqref{mod:outcome}--\eqref{mod:treatment}, we have  $I(U_d;D\mid X)=I(Y_d;D\mid X)$.
\end{lemma}

\begin{proof}
From the model \eqref{mod:outcome}--\eqref{mod:treatment}, namely $Y_d=m_d(X,U_d)$ and $U_d\indep D\mid(Y_d,X)$, it follows that $p(D\mid Y_d,X)=p(D\mid Y_d,U_d, X)=p(D\mid U_d,X)$. Hence, by the definition of conditional mutual information, we obtain
\begin{align*}
 I(Y_d;D\mid X)
% &=\E\left[\log\left(\frac{p(Y_d,D\mid X)}{p(Y_d\mid X)p(D\mid X)}\right)\right]
 =\E\left[\log\left(\frac{p(D\mid Y_d,X)}{p(D\mid X)}\right)\right]
  =\E\left[\log\left(\frac{p(D\mid U_d,X)}{p(D\mid X)}\right)\right]=I(U_d;D\mid X),
 \end{align*}
which proves the result.
\end{proof}

\begin{lemma}\label{lemma:equivalence}
In model  \eqref{mod:outcome}--\eqref{mod:treatment}, we have equivalence between $I(U_d ; D\mid X)\leq I(U_d ; D)$ and $I(U_d;X\mid D)=0$. 
\end{lemma}
\begin{proof} 
By the symmetry of interaction information under permutations, see equation \eqref{eq:perm}, we have
\begin{align*}
 I(U_d;D\mid X)-I(U_d;D) = I(U_d;X\mid D)-I(U_d;X).
  \end{align*}
   The condition $U_d\indep X$,  imposed in model \eqref{mod:outcome}, can be written as $I(U_d;X)=0$ and thus, we obtain
   \begin{align*}
    I(U_d;D\mid X)-I(U_d;D) = I(U_d;X\mid D).
     \end{align*} 
     Since conditional mutual information is nonnegative, $I(U_d;D\mid X)\leq I(U_d;D)$ holds if and only if $I(U_d;X\mid D)=0$.
      \end{proof}

\subsection{Proofs for Section \ref{sec:model}}

\begin{proof}[Proof of Lemma \ref{lem:send:sol}]
 The chain rule for mutual information (see, e.g., \citealp{cover2006}, Theorem 2.5.2) implies
  \begin{align*}
  I(Y_d;D\mid X)+I(X;D)=  I(X;D\mid Y_d)+I(Y_d;D)=I(Y_d;D),
  \end{align*} 
  where the last equality follows from $D\indep X\mid Y_d$, and 
   \begin{align*}
  I(Y_d;D\mid U_d)+I(U_d;D)&= I(U_d;D\mid Y_d)+I(Y_d;D)=I(Y_d;D)
  \end{align*} 
  where the last equality uses $D\indep U_d\mid Y_d$, which follows from $U_d\indep D\mid(Y_d,X)$ and $D\indep X\mid Y_d$.
  Therefore, the inequality $I(Y_d;D\mid X)\leq I(U_d;D)$ as imposed in $\mathrm{CMIR}(d)$ is equivalent to 
  \begin{align*} 
  I(Y_d;D)-I(X;D)\leq I(Y_d;D)-I(Y_d;D\mid U_d), 
  \end{align*} or, equivalently, $I(Y_d;D\mid U_d)\leq I(X;D),$ which proves the result. 
  \end{proof}

\begin{proof}[Proof of Lemma \ref{lem:lat:bal}]
The first claim follows immediately from Lemmas \ref{lem:end} and \ref{lemma:equivalence}. Moreover, when it holds,
\begin{align*}
I(U_d;D\mid X)-I(U_d;D)=I(U_d;X\mid D)=0,
\end{align*}
so that $\mathrm{CMIR}(d)$ holds with equality. It remains to establish the second claim. Let $\varphi$ and $\psi$ be arbitrary bounded measurable functions.  The law of total covariance implies by using $U_d\indep X$ that
\begin{align*}
0&%=\operatorname{Cov}(\varphi(U_d),\psi(X))
=\E\left[\operatorname{Cov}(\varphi(U_d),\psi(X)\mid D)\right]+\operatorname{Cov}\left(\E[\varphi(U_d)\mid D],\E[\psi(X)\mid D]\right)\\
&=\operatorname{Cov}\left(\E[\varphi(U_d)\mid D],\E[\psi(X)\mid D]\right).
\end{align*}
Let $\pi=\P(D=1)$. If $\pi\in\{0,1\}$, the result is immediate. Otherwise, since $D$ is binary,
\begin{align*}
\operatorname{Cov}\left(\E[\varphi(U_d)\mid D],\E[\psi(X)\mid D]\right)&=\pi(1-\pi)\left(\E[\varphi(U_d)\mid D=1]-\E[\varphi(U_d)\mid D=0]\right)\\
&\quad\times\left(\E[\psi(X)\mid D=1]-\E[\psi(X)\mid D=0]\right).
\end{align*}
Hence, for every pair of bounded measurable functions $\varphi$ and $\psi$,
\begin{align*}
\left(\E[\varphi(U_d)\mid D=1]-\E[\varphi(U_d)\mid D=0]\right)\left(\E[\psi(X)\mid D=1]-\E[\psi(X)\mid D=0]\right)=0.
\end{align*}
If the first factor is zero for every $\varphi$, then $U_d\indep D$. Otherwise, there exists a $\varphi$ for which the first factor is nonzero, in which case the second factor must be zero for every $\psi$, implying $X\indep D$.
\end{proof}

\subsection{Proofs for Section \ref{sec:im}}

\begin{proof}[Proof of Theorem \ref{thm:mult}]
By Lemma \ref{lem:end} and \ref{lemma:equivalence}, the CMI restriction $I(Y_d;D\mid X)\leq I(U_d;D)$ is equivalent to $I(U_d;X\mid D)=0$. Since $D$ is discrete, we have
\begin{align*}
I(U_d;X\mid D)
=
\sum_{d'=0}^K \P(D=d') I(U_d;X\mid D=d').
\end{align*}
Since $\mathcal D=\operatorname{supp}(D)$, we have $\P(D=d')>0$ for every $d'\in\mathcal D$. Hence, by nonnegativity of mutual information, $I(U_d;X\mid D=d')=0$, and thus $U_d\indep X\mid D=d'$ for every $d'$.

It remains to show that this cellwise conditional independence is equivalent to the multiplicative representation. For any $d'\in\mathcal D$, Bayes' rule, together with $U_d\indep X$ and $U_d\indep X\mid D=d'$, gives
\begin{align*}
\P(D=d'\mid U_d,X)
=
\frac{\P(D=d'\mid U_d)\P(D=d'\mid X)}
{\P(D=d')}.
\end{align*}
Hence, the desired multiplicative representation follows. Conversely, suppose that
\begin{align*}
\P(D=d'\mid U_d,X)
=
g_{d,d'}(U_d)h_{d'}(X).
\end{align*}
Since $U_d\indep X$, the conditional distribution of $(U_d,X)$ given $D=d'$ then factorizes into its conditional marginals, so that
$U_d\indep X\mid D=d'$. Indeed, for arbitrary bounded measurable functions $\varphi$ and $\psi$, the multiplicative representation and $U_d\indep X$ imply
\begin{align*}
\E[\varphi(U_d)\psi(X)\mid D=d']
=
\E[\varphi(U_d)\mid D=d']\,
\E[\psi(X)\mid D=d'],
\end{align*} 
Since $d'$ was arbitrary, the multiplicative representation for all $d'\in\mathcal D$ is equivalent to
$U_d\indep X\mid D$, and hence to $\mathrm{CMIR}(d)$.
\end{proof}

\begin{proof}[Proof of Corollary \ref{coro:mult:d}]
Since $Y_d=m_d(X,U_d)$ and $U_d\indep D\mid(Y_d,X)$, we have $\P(D=d'\mid U_d,X)=\P(D=d'\mid Y_d,X)$. Taking conditional expectations of the corresponding log likelihood ratios given $D=d'$ yields $I_{d'}(U_d;D\mid X)=I_{d'}(Y_d;D\mid X)$.
Moreover, using $U_d\indep X$ we obtain
\begin{align*}
I_{d'}(U_d;D\mid X)-I_{d'}(U_d;D)
= \E\left[\left.\log\left(\frac{p(U_d\mid D=d',X)} {p(U_d\mid D=d')}\right)\right|D=d'\right] =I(U_d;X\mid D=d')\geq 0.
\end{align*}
Hence, $I_{d'}(Y_d;D\mid X)\leq I_{d'}(U_d;D)$ is equivalent to $U_d\indep X\mid D=d'$. The result thus follows by adapting the proof of Theorem \ref{thm:mult} to the $d'$-specific case. 
\end{proof}

\begin{proof}[Proof of Theorem \ref{thm:uc:mult}]
If $Y_d\indep D\mid X$, then $\mathrm{CMIR}(d)$ follows immediately from $I(Y_d;D\mid X)=0$. For the converse, Theorem \ref{thm:mult} implies that $\mathrm{CMIR}(d)$ is equivalent to the existence of nonnegative measurable
functions $g_{d,d'}$ and $h_{d'}$ such that
\begin{align*}
\P(D=d'\mid U_d,X)=g_{d,d'}(U_d)h_{d'}(X)
\end{align*}
Since $U_d\indep X$, iterated expectations imply
\begin{align*}
\P(D=d'\mid X)=h_{d'}(X)\E[g_{d,d'}(U_d)]
\end{align*}
Assumption \ref{A:mult:relevance} implies that $\P(D=d'\mid X)>0$ on a set of positive probability. Since $\P(D=d'\mid X)=h_{d'}(X)\E[g_{d,d'}(U_d)]$, with $h_{d'}$ finite and nonnegative, it follows that $0<\E[g_{d,d'}(U_d)]<\infty$.
Define $\widetilde g_{d,d'}(u)=g_{d,d'}(u)/\E[g_{d,d'}(U_d)]$. Then $\E[\widetilde g_{d,d'}(U_d)]=1$ and
\begin{align*}
\P(D=d'\mid U_d,X)=\widetilde g_{d,d'}(U_d)\P(D=d'\mid X)
\end{align*}

Since the conditional treatment probabilities sum to one, the normalized factorization implies, using the notation $p(X)=(\P(D=0\mid X),\ldots,\P(D=K\mid X))^\top$, that
\begin{align*}
\E\left[p(X)p(X)^\top\right]\bigl(\widetilde g_{d,0}(u)-1,\ldots,\widetilde g_{d,K}(u)-1\bigr)^\top=0
\end{align*}
for almost all $u$.
Assumption \ref{A:mult:relevance}, i.e., that $\E[p(X)p(X)^\top]$ is nonsingular, therefore implies
$\widetilde g_{d,d'}(u)=1$ for every $d'\in\mathcal D$ and almost every $u$. Consequently, $\P(D=d'\mid U_d,X)=\P(D=d'\mid X)$ for every $d'\in\mathcal D$, and therefore $D\indep U_d\mid X$. Since $Y_d=m_d(X,U_d)$, it follows that $Y_d\indep D\mid X$.
\end{proof}

\begin{proof}[Proof of Corollary \ref{thm:uc}]
It is sufficient to prove the converse direction and thus suppose that $\mathrm{CMIR}(d)$ holds. From Theorem \ref{thm:mult}, there exist nonnegative measurable functions $g_{d,d'}$ and $h_{d'}$ such that
\begin{align*}
\P(D=d'\mid U_d,X)=g_{d,d'}(U_d)h_{d'}(X), \qquad d'\in\{0,1\}.
\end{align*}
Since the two conditional treatment probabilities sum to one,
\begin{align*}
1-g_{d,1}(U_d)h_1(X)=g_{d,0}(U_d)h_0(X)
\end{align*}
Since $U_d\indep X$, this equality holds for almost every $(u,x)$ with respect to the product distribution of $(U_d,X)$. Hence, by Fubini's theorem, for almost all $(u_1,u_2,x_1,x_2)$, it holds simultaneously at $(u_1,x_1)$, $(u_2,x_2)$, $(u_1,x_2)$, and $(u_2,x_1)$. Multiplying crosswise and using that the resulting right-hand sides coincide yields
\begin{align*}
&\bigl(1-g_{d,1}(u_1)h_1(x_1)\bigr)
\bigl(1-g_{d,1}(u_2)h_1(x_2)\bigr)=\bigl(1-g_{d,1}(u_1)h_1(x_2)\bigr)
\bigl(1-g_{d,1}(u_2)h_1(x_1)\bigr).
\end{align*}
Expanding and cancelling terms gives
\begin{align}\label{eq:key_id}
\bigl(g_{d,1}(u_1)-g_{d,1}(u_2)\bigr)
\bigl(h_1(x_1)-h_1(x_2)\bigr)
=
0
\end{align}
for almost all $(u_1,u_2,x_1,x_2)$.

Since $U_d\indep X$, iterated expectations and the multiplicative factorization imply
\begin{align*}
\P(D=1\mid X)
=
\E[g_{d,1}(U_d)]h_1(X).
\end{align*}
By assumption, $\P(D=1\mid X)$ is not almost surely constant. Hence, $h_1(X)$ is not almost surely constant. In particular, for independent copies $X_1$ and $X_2$ of $X$,
\begin{align*}
\P\bigl(h_1(X_1)\neq h_1(X_2)\bigr)>0.
\end{align*}
Together with \eqref{eq:key_id}, this implies
$g_{d,1}(u_1)=g_{d,1}(u_2)$ for almost all $(u_1,u_2)$, so that $g_{d,1}$ is constant almost surely. Another application of iterated expectations therefore gives $\P(D=1\mid U_d,X)=\P(D=1\mid X)$. Since $D$ is binary, this implies $D\indep U_d\mid X$, and thus the result follows.
\end{proof}

\begin{proof}[Proof of Theorem \ref{thm:uc:state}]
Suppose that $\mathrm{CMIR}(d',d)$ holds. By Corollary \ref{coro:mult:d} and $U_d\indep X$, the normalized factorization is
\begin{align}\label{eq:normalized-factorization}
\P(D=d'\mid U_d,X)
=
\widetilde g_{d,d'}(U_d)\P(D=d'\mid X),
\qquad
\E[\widetilde g_{d,d'}(U_d)]=1.
\end{align}
Since the left-hand side is a probability,
\begin{align}\label{eq:probability-bound}
\widetilde g_{d,d'}(u)\P(D=d'\mid X)\leq1
\end{align}
for almost every $u$.  If $\widetilde g_{d,d'}(u)>1/\kappa$ for such a $u$, the support condition with $\varepsilon=\kappa-1/\widetilde g_{d,d'}(u)>0$ implies
\begin{align*}
\P\left(\P(D=d'\mid X)>\frac{1}{\widetilde g_{d,d'}(u)}\right)>0,
\end{align*}
contradicting \eqref{eq:probability-bound}. Hence, $\widetilde g_{d,d'}(U_d)\leq1/\kappa$ almost surely.

Integrating \eqref{eq:normalized-factorization} over $X$ and using $U_d\indep X$ gives $\P(D=d'\mid U_d)
=
\widetilde g_{d,d'}(U_d)\P(D=d')$. By Bayes' rule, $\widetilde g_{d,d'}$ is therefore the density ratio of the conditional distribution of $U_d$ given $D=d'$ relative to its marginal distribution. Consequently,
\begin{align*}
I_{d'}(U_d;D)
&=\E\left[\left.\log\widetilde g_{d,d'}(U_d)\right|D=d'\right]\leq\log(1/\kappa).
\end{align*}
Under $\mathrm{CMIR}(d',d)$, the proof of Corollary \ref{coro:mult:d} gives $I_{d'}(Y_d;D\mid X)=I_{d'}(U_d;D)$. Combining this equality with nonnegativity and the preceding bound yields
\begin{align*}
0\leq I_{d'}(Y_d;D\mid X)=I_{d'}(U_d;D)\leq\log(1/\kappa).
\end{align*}

If $\kappa=1$, the bound $\widetilde g_{d,d'}(U_d)\leq1$ and the normalization $\E[\widetilde g_{d,d'}(U_d)]=1$ imply $\widetilde g_{d,d'}(U_d)=1$ almost surely. Thus, $\P(D=d'\mid U_d,X)=\P(D=d'\mid X)$. Conditioning on $(Y_d,X)$ gives $\P(D=d'\mid Y_d,X)=\P(D=d'\mid X)$. Conversely, this equality implies $I_{d'}(Y_d;D\mid X)=0$, so $\mathrm{CMIR}(d',d)$ follows from the nonnegativity of $I_{d'}(U_d;D)$.
\end{proof}

\subsection{Proofs for Section \ref{sec:ext}}\label{appendix:ext}
The following two corollaries are the conditional counterparts of Lemmas \ref{lem:end} and \ref{lemma:equivalence}, respectively. Both follow by additionally conditioning on $W$; we include their short proofs for completeness.
\begin{corollary}\label{coro:end:cond}
In model \eqref{mod:outcome:cond}--\eqref{mod:treatment:cond}, we have $I(U_d;D\mid X,W)=I(Y_d;D\mid X,W)$.
\end{corollary}

\begin{proof}
By the conditional chain rule for mutual information and $Y_d=m_d(X,W,U_d)$, we have $I(U_d;D\mid X,W)=I(Y_d;D\mid X,W)+I(U_d;D\mid Y_d,X,W)$. By the maintained assumption $U_d\indep D\mid(Y_d,X,W)$, the second term is zero, which gives the result.
\end{proof}

\begin{corollary}[Conditional equivalence]\label{coro:equivalence:cond}
Suppose that $U_d\indep X\mid W$. Then $I(U_d;D\mid X,W)\leq I(U_d;D\mid W)$
is equivalent to $I(U_d;X\mid D,W)=0$.
\end{corollary}
\begin{proof}
By the conditional interaction-information identity,
\begin{align*}
I(U_d;D\mid W)-I(U_d;D\mid X,W)=I(U_d;X\mid W)-I(U_d;X\mid D,W).
\end{align*}
Hence, $I(U_d;D\mid X,W)\leq I(U_d;D\mid W)$ is equivalent to $I(U_d;X\mid D,W)\leq I(U_d;X\mid W)$. The result follows from the maintained assumption $U_d\indep X\mid W$ and the nonnegativity of conditional mutual information.
\end{proof}

\begin{proof}[Proof of Theorem \ref{thm:mult:cond}]
By Corollaries \ref{coro:equivalence:cond} and \ref{coro:end:cond}, $\mathrm{CMIR}_W(d)$ is equivalent to $U_d\indep X\mid(D,W)$. It therefore remains to establish the equivalence between this conditional independence and the multiplicative representation.

The forward implication follows from the same Bayes argument as in the proof of Theorem \ref{thm:mult}, applied conditional on $W$. In particular, for almost all $w$ such that $\P(D=d'\mid W=w)>0$,
\begin{align*}
\P(D=d'\mid U_d=u,X=x,W=w)=\frac{\P(D=d'\mid U_d=u,W=w)\P(D=d'\mid X=x,W=w)}{\P(D=d'\mid W=w)}
\end{align*}
for almost all $(u,x)$, where we use $U_d\indep X\mid W$ and $U_d\indep X\mid(D=d',W)$. Hence, the desired factorization follows by setting
\begin{align*}
g_{d,d'}(u,w)&=\frac{\P(D=d'\mid U_d=u,W=w)}{\P(D=d'\mid W=w)}, \qquad h_{d'}(x,w)=\P(D=d'\mid X=x,W=w).
\end{align*}
If $\P(D=d'\mid W=w)=0$, the factorization can be extended by setting $g_{d,d'}(u,w)=1$ and $h_{d'}(x,w)=0$.

Conversely, suppose that $\P(D=d'\mid U_d,X,W)=g_{d,d'}(U_d,W)h_{d'}(X,W)$. For almost every $w$ such that $\P(D=d'\mid W=w)>0$, define $G_{d,d'}(w)=\E[g_{d,d'}(U_d,w)\mid W=w]$ and $H_{d'}(w)=\E[h_{d'}(X,w)\mid W=w]$. Using $U_d\indep X\mid W$ and taking conditional expectations yields
\begin{align*}
\P(D=d'\mid W=w)=G_{d,d'}(w)H_{d'}(w)>0.
\end{align*}
Both factors are therefore finite and strictly positive. For arbitrary bounded measurable functions $\varphi$ and $\psi$, the factorization and $U_d\indep X\mid W$ imply
\begin{align*}
\E[\varphi(U_d)\psi(X)\mathbbm{1}_{\{D=d'\}}\mid W=w]&=\E\!\left[\varphi(U_d)\psi(X)\P(D=d'\mid U_d,X,W)\mid W=w\right]\\
&=\E\!\left[\varphi(U_d)g_{d,d'}(U_d,w)\psi(X)h_{d'}(X,w)\mid W=w\right]\\
&=\E[\varphi(U_d)g_{d,d'}(U_d,w)\mid W=w]\,\E[\psi(X)h_{d'}(X,w)\mid W=w].
\end{align*}
Whenever $\P(D=d'\mid W=w)>0$, dividing by $\P(D=d'\mid W=w)=G_{d,d'}(w)H_{d'}(w)$ gives
\begin{align*}
\E[\varphi(U_d)\psi(X)\mid D=d',W=w]&=\frac{\E[\varphi(U_d)g_{d,d'}(U_d,w)\mid W=w]}{G_{d,d'}(w)}\frac{\E[\psi(X)h_{d'}(X,w)\mid W=w]}{H_{d'}(w)}\\
&=\E[\varphi(U_d)\mid D=d',W=w]\E[\psi(X)\mid D=d',W=w].
\end{align*}
Hence, $U_d\indep X\mid(D=d',W)$. Since $d'$ was arbitrary, we obtain $U_d\indep X\mid(D,W)$, which completes the proof.
\end{proof}

\begin{proof}[Proof of Corollary \ref{coro:uc:mult:cont}]
If $Y_d\indep D\mid X$, then $\mathrm{CMIR}(d)$ follows immediately from $I(Y_d;D\mid X)=0$. For the converse, suppose that $\mathrm{CMIR}(d)$ holds. By the continuous-treatment analogue of Theorem \ref{thm:mult}, we have
\begin{align*}
f_{D\mid U_d,X}(t\mid u,x)=g_d(t,u)h(t,x)
\end{align*}
for almost all $(t,u,x)$. Since $U_d\indep X$, integrating with respect to $U_d$ gives
\begin{align*}
f_{D\mid X}(t\mid x)=\E[g_d(t,U_d)]h(t,x).
\end{align*}
Recall that $\widetilde g_d(t,u)=g_d(t,u)/\E[g_d(t,U_d)]$ whenever $0<\E[g_d(t,U_d)]<\infty$, and $\widetilde g_d(t,u)=1$ otherwise.
The normalized factorization then gives
\begin{align*}
f_{D\mid U_d,X}(t\mid u,x)=\widetilde g_d(t,u)f_{D\mid X}(t\mid x)
\end{align*}
for almost every $(t,u,x)$. Hence, for almost every $u$,
\begin{align*}
\E[\widetilde g_d(D,u)-1\mid X=x]
&=\int_{\mathcal D}\left(\widetilde g_d(t,u)-1\right)f_{D\mid X}(t\mid x)\,\mathrm dt\\
&=\int_{\mathcal D}\left(f_{D\mid U_d,X}(t\mid u,x)-f_{D\mid X}(t\mid x)\right)\,\mathrm dt=0.
\end{align*}
Nonnegativity and normalization imply $\E[\widetilde g_d(D,u)]=1$ and hence $\E[|\widetilde g_d(D,u)-1|]\leq2$ for almost every $u$. Assumption \ref{A:cont:relevance} therefore implies $\widetilde g_d(D,u)=1$ almost surely for almost every $u$. Consequently, $f_{D\mid U_d,X}(t\mid u,x)=f_{D\mid X}(t\mid x)$ almost everywhere, so that $U_d\indep D\mid X$. Since $Y_d=m_d(X,U_d)$, the result follows.
\end{proof}

\begin{proof}[Proof of Theorem \ref{thm:apo:pointwise}]
From the proof of Corollary \ref{coro:uc:mult:cont}, we know that $\widetilde g_d(\cdot,u)-1\in\mathcal N(T)$ for almost every $u$. Since $\widetilde g_d(\cdot,u)-1\in\mathcal H$, Assumption \ref{A:pointwise:relevance} implies $\widetilde g_d(d,u)=1$ for almost every $u$. Moreover, $\mathrm{CMIR}(d)$ implies the normalized factorization
\begin{align*}
f_{D\mid U_d,X}(t\mid u,x)=\widetilde g_d(t,u)f_{D\mid X}(t\mid x)
\end{align*}
almost everywhere.

For almost every $x$, continuity and positivity of $f_{D\mid X}(\cdot\mid x)$ at $d$ imply positivity in a neighborhood of $d$. Bayes' rule, the factorization, and $U_d\indep X$ therefore imply that, for almost every $t$ in this neighborhood, the conditional density of $U_d$ given $(D=t,X=x)$ relative to its marginal distribution is $\widetilde g_d(t,\cdot)$. In particular, $\E[\widetilde g_d(t,U_d)]=1$. By continuity at $d$, $\widetilde g_d(t,U_d)\to1$ almost surely. Scheffé's lemma consequently yields $\E[|\widetilde g_d(t,U_d)-1|]\to 0$ as $t\to d$ through these treatment values.

Let $M<\infty$ satisfy $|Y_d|\leq M$ almost surely. Since $Y_d=m_d(X,U_d)$ and $U_d\indep X$, it follows that, for almost every $x$,
\begin{align*}
\left|\E[Y_d\mid D=t,X=x]-\E[Y_d\mid X=x]\right|
&=\left|\E\left[m_d(x,U_d)\bigl(\widetilde g_d(t,U_d)-1\bigr)\right]\right|\\
&\leq M\E[|\widetilde g_d(t,U_d)-1|]
\to 0.
\end{align*}
Furthermore, consistency, $Y=Y_D$, and Assumption \ref{A:outcome:lipschitz} imply
\begin{align*}
\left|\E[Y\mid D=t,X=x]-\E[Y_d\mid D=t,X=x]\right|
\leq \E\big[|Y_D-Y_d|\bigm| D=t,X=x\big]\leq L\|t-d\|
\end{align*}
for almost all $t\in\mathcal D$ satisfying $\|t-d\|<\delta$.
Combining these bounds gives $\E[Y\mid D=t,X=x]\to\E[Y_d\mid X=x]$ for almost every $x$, as $t\to d$ outside a set of treatment values of Lebesgue measure zero. Defining the conditional regression at $d$ by this limit establishes the claimed equality.
\end{proof}

\section{Testing Covariate Relevance: Simulations}\label{sec:simulation}

We examine the finite-sample performance of the rank test of \citet{ChenFang2019} for assessing multinomial covariate relevance in Assumption \ref{A:mult:relevance}. We consider discrete and continuous covariates, using $5{,}000$ Monte Carlo replications, a nominal significance level of $5\%$, and sample sizes $n\in\{500,1{,}000,5{,}000\}$.

Let $D\in\{0,1,2\}$ and let $U_d$ be standard normal and independent of $X$. Treatment combines the latent-selection mechanism of Example \ref{ex:within} with assignment based only on covariates. For $\theta\in[0,1]$, define $w_\theta=\min\{1,\max\{0,3\theta-1\}\}$ and $q_{j,\theta}(X)=(1-\theta^2)/3+\theta^2q_j(X)$. The conditional treatment probabilities are
\begin{align*}
\P(D=0\mid U_d,X)&=(1-w_\theta)\Phi(\theta-U_d)+w_\theta q_{0,\theta}(X),\\
\P(D=1\mid U_d,X)&=(1-w_\theta)\Phi(U_d-\theta)\bigl(1-a(X)\bigr)+w_\theta q_{1,\theta}(X),\\
\P(D=2\mid U_d,X)&=(1-w_\theta)\Phi(U_d-\theta)a(X)+w_\theta q_{2,\theta}(X),
\end{align*}
where $\Phi$ denotes the standard normal distribution function. 
For discrete covariates, $X$ takes values $-1,0,1$ with equal probabilities, and $a(-1)=1/4$, $a(0)=1/2$, and $a(1)=3/4$. The probabilities $q_j(X)$ are the rows of
\[
Q=\begin{pmatrix}
3/4&1/8&1/8\\
1/8&3/4&1/8\\
1/8&1/8&3/4
\end{pmatrix},
\]
where columns correspond to treatment states. For continuous covariates, $X$ is standard normal, $a(X)=\Phi(X)$, and $(q_0(X),q_1(X),q_2(X))=(\Phi(X)^2,1-\Phi(X),\Phi(X)(1-\Phi(X)))$.

Let $p(X)$ denote the propensity-score vector and $G=\E[p(X)p(X)^\top]$. For $\theta\leq1/3$, CMIR holds and $G$ has rank two, while changes in $\theta$ shift the participation threshold. For $1/3<\theta<2/3$, $G$ has full rank but treatment remains related to latent heterogeneity. Hence, CMIR fails for the potential outcome model $Y_d=\alpha_d+\beta_dX+U_d$. For $\theta\geq2/3$, treatment depends only on covariates, so unconfoundedness and CMIR hold. Within this interval, increasing $\theta$ strengthens covariate relevance.

\paragraph{Implementation.}
We test $H_0:\operatorname{rank}(G)\leq2$ against full rank using $M=\E[b(X)T^\top]$, where $T=(\mathbbm{1}_{\{D=0\}},\mathbbm{1}_{\{D=1\}},\mathbbm{1}_{\{D=2\}})^\top$. For discrete covariates, $b(X)$ consists of the three covariate-group indicators. For continuous covariates, we use $b(X)=(1,\sqrt{3}(2\Phi(X)-1),\sqrt{5}(6\Phi(X)^2-6\Phi(X)+1))^\top$. In both designs, this basis spans the propensity scores and $\E[b(X)b(X)^\top]$ is positive definite, implying $\operatorname{rank}(M)=\operatorname{rank}(G)$. We apply the analytic bootstrap procedure of \citet{ChenFang2019} to $\widehat M=n^{-1}\sum_i b(X_i)T_i^\top$, estimating the null rank using the singular-value threshold $n^{-1/4}\sigma_1(\widehat M)$ and obtaining critical values from $1{,}000$ Gaussian multiplier bootstrap draws.
\begin{figure}[htbp]
\centering
\includegraphics[width=1\textwidth]{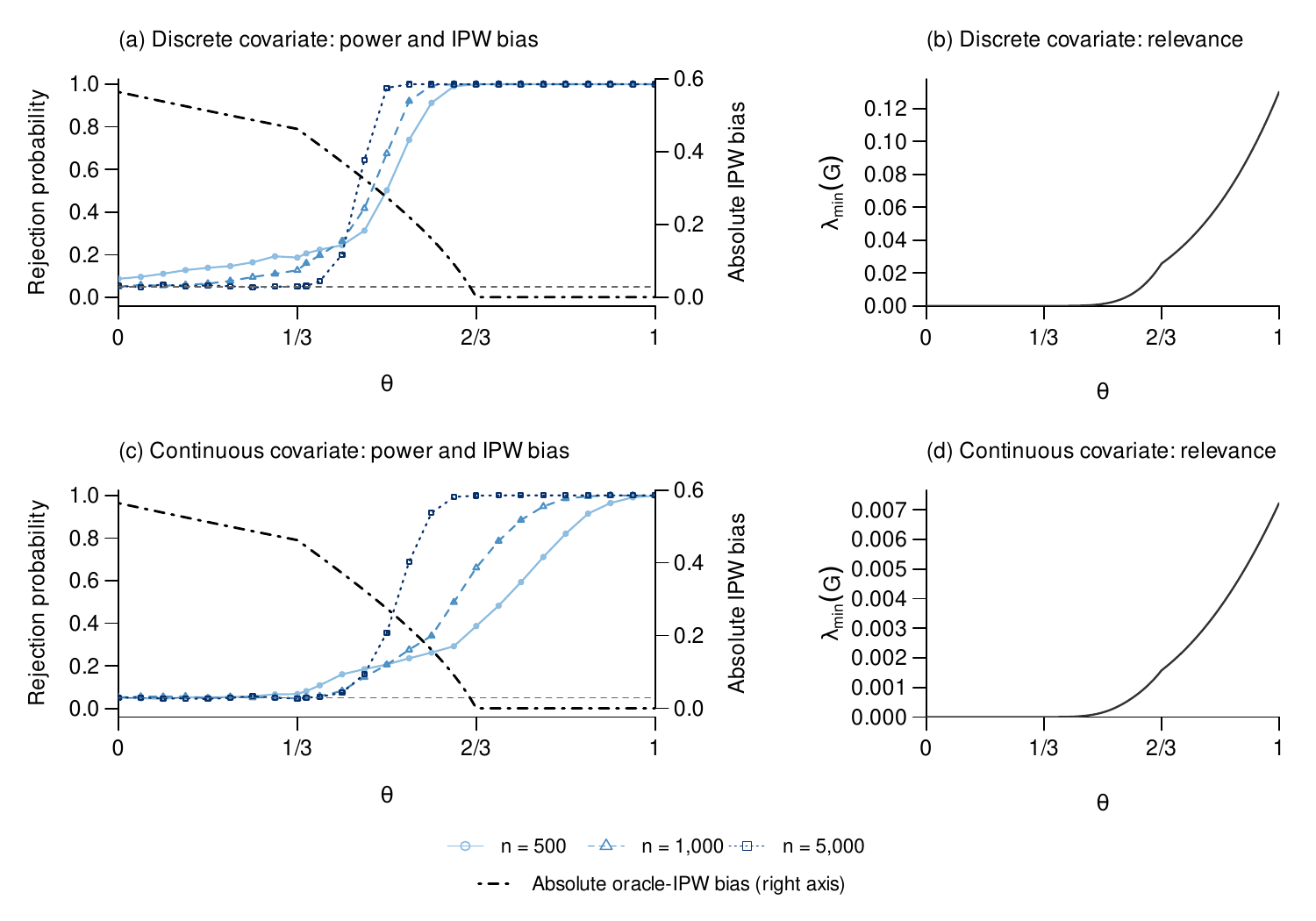}
\caption{Rejection probabilities of the Chen--Fang rank test and absolute oracle-IPW bias (left), and minimum eigenvalues of $G$ (right). The null holds for $\theta\leq1/3$; CMIR holds on $[0,1/3]$ and $[2/3,1]$. Horizontal dashed lines mark the $5\%$ nominal level. }
\label{fig:rank:power}
\end{figure}

We also consider oracle IPW estimation of $\E[Y_0]$, where $Y_0=\alpha_0+\beta_0X+U_0$ and $U_0=U_d$ in the assignment mechanism. The true control propensity is $p_0(X)=(1-w_\theta)\Phi(\theta/\sqrt{2})+w_\theta q_{0,\theta}(X)$. For $\theta<2/3$, the estimator $\widehat\mu_0^{\mathrm{IPW}}=n^{-1}\sum_i\mathbbm{1}_{\{D_i=0\}}Y_i/p_0(X_i)$ has population bias
\[
\E[\widehat\mu_0^{\mathrm{IPW}}]-\E[Y_0]
=-\frac{(1-w_\theta)e^{-\theta^2/4}}{2\sqrt{\pi}}\E\left[\frac{1}{p_0(X)}\right],
\]
using $U_0\indep X$, $\E[U_0]=0$, and $\E[U_0\Phi(U_0-\theta)]=e^{-\theta^2/4}/(2\sqrt{\pi})$. The bias vanishes for $\theta\geq2/3$. We plot its absolute value, which does not depend on sample size.

\paragraph{Size and power.}
The null holds for $\theta\leq1/3$, while $\theta>1/3$ generates full-rank alternatives. In the discrete design, null rejection frequencies range from $0.087$ to $0.193$ at $n=500$, indicating overrejection. 
The corresponding ranges are $0.050$--$0.127$ at $n=1{,}000$ and $0.047$--$0.061$ at $n=5{,}000$. In the continuous design, null rejection frequencies remain between $0.045$ and $0.068$ across all sample sizes. Thus, size accuracy improves with sample size in the discrete design and is generally better in the continuous design.

Figure \ref{fig:rank:power} shows that power generally increases with covariate relevance and sample size. At $\theta=0.5$, rejection frequencies are $0.502$, $0.675$, and $0.983$ in the discrete design, compared with $0.207$, $0.205$, and $0.356$ in the continuous design. The more rapid increase in discrete-design power reflects differences in the assignment mechanisms rather than an inherent advantage of discrete covariates.

Substantial oracle-IPW bias can coexist with high power against rank deficiency. In the intermediate interval, full rank holds but CMIR fails, illustrating why covariate relevance alone does not justify unconfoundedness. Since the estimator uses the true propensity scores, its bias reflects residual confounding rather than propensity-score estimation error.

\bibliographystyle{econometrica}

\bibliography{references}

\end{document}